\documentclass[12pt,a4paper]{article}
\usepackage[top=2.5cm, bottom=2.5cm, left=2.5cm, right=2.5cm]{geometry}
\usepackage[english]{babel}

\usepackage{mathptmx} 
\usepackage[T1]{fontenc}
\usepackage[utf8]{inputenc}

\usepackage{helvet}
\usepackage{xcolor}
\usepackage{amsmath,amssymb}

\usepackage[super]{nth}

\usepackage[absolute]{textpos}
\usepackage{atbegshi}
\usepackage{graphicx}
\usepackage{subcaption}

\graphicspath{{figures/}{Images/}}

\usepackage{parskip}
\usepackage{setspace}
\usepackage[export]{adjustbox}

\usepackage{titlesec}
\titleformat{\section}[block]{\normalfont\bfseries}{\thesection}{1em}{\MakeUppercase}
\titleformat{\subsection}[block]{\normalfont\bfseries}{\thesubsection}{0.5em}{}
\titlespacing*{\section}{0pt}{8pt plus 2pt minus 2pt}{4pt plus 1pt}
\titlespacing*{\subsection}{0pt}{6pt plus 2pt minus 2pt}{3pt plus 1pt}

\usepackage{fancyhdr}
\AtBeginDocument{%
    \lfoot{\vspace*{-1.1cm}\printemail}%
    \rfoot{}
    \AtBeginShipoutNext{%
        \lfoot{}%
        \rfoot{\vspace*{-0.8cm}\thepage}
    }%
}
\usepackage{comment}
\usepackage{booktabs}
\usepackage{tabularx}
\usepackage{multirow}
\usepackage{float}
\usepackage{blkarray}
\usepackage{enumitem}

\usepackage{cite}
\usepackage[bookmarks=false,hidelinks]{hyperref}
\hypersetup{%
    plainpages=false,%
    pdfsubject=CIGRE Grid of the Future 2026,%
    pdfstartview=FitH%
}

\usepackage{tikz}
\usetikzlibrary{calc,positioning,arrows.meta,shapes.geometric,fit,backgrounds}
\definecolor{CIGgreen}{RGB}{0,91,0}

\usepackage{multicol}
\usepackage{xcolor,colortbl}
\usepackage{floatflt}
\usepackage{wrapfig}

\usepackage[font={small,it}]{caption}

\newcommand{\PS}{\textsc{ps}}
\newcommand{\LC}{\textsc{lc}}
\newcommand{\BA}{\textsc{ba}}
\newcommand{\CI}{\textsc{ci}}
\newcommand{\AR}{\textsc{ar}}
\newcommand{\Scal}{\mathcal{S}}

\newcommand{\Nbusy}{N_{\mathrm{busy}}}

\makeatletter
\def\printtitle{%
    \fontsize{12}{14.5}{
        {{\textbf{\textsf{\@paperTitle\\}}}}}}
\def\printauthor{%
    \fontsize{12}{14.5}{
        {{\textbf{\@authorName \\ \@authorCompany \\ \@authorCountry}}}}}
\def\printpaperRef{%
    \textsf{\color{CIGgreen}
        {\@paperRef}}}
\def\printemail{%
    \textbf{{\@authorEmail}}}

\newcommand{\authorName}[1]{\gdef\@authorName{#1}}
\newcommand{\@authorName}{\@latex@error{No \noexpand\authorName given}\@ehc}

\newcommand{\authorCompany}[1]{\gdef\@authorCompany{#1}}
\newcommand{\@authorCompany}{\@latex@error{No \noexpand\authorCompany given}\@ehc}

\newcommand{\authorCountry}[1]{\gdef\@authorCountry{#1}}
\newcommand{\@authorCountry}{\@latex@error{No \noexpand\authorCountry given}\@ehc}

\newcommand{\paperRef}[1]{\gdef\@paperRef{#1}}
\newcommand{\@paperRef}{\@latex@error{No \noexpand\paperRef given}\@ehc}

\newcommand{\paperTitle}[1]{\gdef\@paperTitle{#1}}
\newcommand{\@paperTitle}{\@latex@error{No \noexpand\paperTitle given}\@ehc}

\newcommand{\authorEmail}[1]{\gdef\@authorEmail{#1}}
\newcommand{\@authorEmail}{\@latex@error{No \noexpand\authorEmail given}\@ehc}

\makeatother

\newcommand{\CIGREtitle}{
    \begin{textblock*}{\textwidth}(23mm,14.8mm)
        \includegraphics[width=2.95cm]{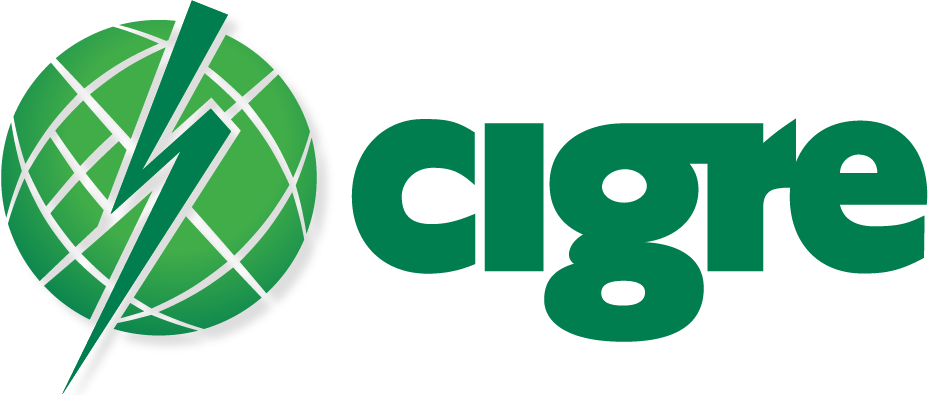}
    \end{textblock*}

    \begin{textblock*}{\textwidth}(25mm,40mm)
        \centering
        \fontsize{12}{12}{
            \textsf{\printpaperRef}}
    \end{textblock*}

    \begin{textblock*}{\textwidth}(25mm,50mm)
        \centering
        \printtitle
        \vspace{12pt}
        \printauthor
    \end{textblock*}
    \vspace*{8.0cm} 
}

\paperTitle{A Hierarchical Semi-Markov Load Model for AI Data Centers Coupling Job Scheduling with Bulk-Synchronous-Parallel Power Dynamics}

\authorName{\textbf{\fontsize{12pt}{14pt}\selectfont
\underline{C}.~CHAUDHARY$^{1}$,
\underline{A}.~BERA$^{2}$,
\underline{C}.~NEWLUN$^{2}$,
\underline{M}.~BENIDRIS$^{1}$,
\underline{J}.~MITRA$^{1}$}
\vspace*{4pt}}

\authorCompany{$^{1}$Michigan State University, East Lansing, MI 48824 \\
$^{2}$Sandia National Laboratories, Albuquerque, NM 87185}

\authorCountry{United States of America}

\authorEmail{chaud152@msu.edu}

\paperRef{}

\begin{document}
    \CIGREtitle 

    \sloppy
    \vspace{-1cm}
    \section*{Summary}
    AI data centers are emerging as a dominant new load class on the bulk power system, and their power dynamics differ fundamentally from those of conventional industrial loads. Within a single training job, the bulk-synchronous-parallel (BSP) algorithm drives every node through a cycle of dense computation, synchronization, and checkpoint writes, which produces power swings between thermal design power and near idle within seconds. Across the facility, training jobs arrive, claim blocks of nodes for hours to days, and depart, so occupied-node counts fluctuate daily, weekly, and yearly. The facility-scale variability and the peak demand that sizes the grid interconnection are governed by this second, slower mechanism. A model that resolves only the within-job physics and treats the facility as a fixed population of busy nodes averages out the per-node swings and underestimates the peak-to-average ratio.

This paper develops a hierarchical semi-Markov Data-Center (HSM-DC) load model that couples two layers across two timescales. A job-scheduling layer generates training jobs through a non-homogeneous compound-Poisson arrival process shaped by diurnal, weekly, and seasonal patterns, assigns heavy-tailed node counts and durations, and places jobs on a finite node pool through a first-in-first-out scheduler. A within-job layer drives each occupied node with a five-state semi-Markov chain covering the phases of a BSP iteration, with state-dependent Ornstein-Uhlenbeck noise. IT power follows from the fluctuating node occupancy and the per-node power, and the busy-node power, the idle floor, and the saturation ceiling are anchored to measured node telemetry and to the straight-line facility-power-versus-utilization relationship of a whole-facility reference.

Configured to the reference colocation facility at matched scale, the model reproduces the whole-facility mean IT power, standard deviation, and peak-to-average ratio across utilization with coefficients of determination of 0.9997, 0.92, and 0.82, respectively. It also reproduces the queued-job fraction to within one percentage point at high utilization. The operational conclusion is direct. Facility-scale variability and peak demand derive from the job-arrival and scheduling process, so interconnection planning must represent that process instead of scaling a node-level power profile.

    \section*{Keywords}
    AI data centers, AI workload profile, bulk-synchronous-parallel training, data center load model, job scheduling, peak-to-average ratio, semi-Markov load model.

    \pagebreak 

\pagestyle{empty}

    \section{Introduction}
    AI data centers exemplify a new class of large load whose demand pattern departs from the smooth, slowly varying profile of conventional large loads, such as steady industrial plants and other base-load facilities that do not exhibit the rapid, structured power swings addressed here, and interconnection and stability studies have not kept pace with this shift. United States data center consumption rose from roughly 76~TWh in 2018 to about 176~TWh in 2023, near 4.4\% of national electricity use, and is projected to double or triple again before the end of the decade as generative-AI training and inference scale~\cite{sheng2026power,shehabi2024lbnl}. At the campus level, this growth appears as hundred-megawatt interconnection requests that rival large industrial loads and, in some regions, exceed local transmission capacity, and system operators have already documented megawatt-scale load steps and the abrupt loss of more than a gigawatt of demand within seconds following a transmission fault~\cite{nerc2025largeloads}.

This growth carries direct consequences for planning. Interconnection studies size transmission and generation capacity from a load's expected peak and its peak-to-average ratio (PAR), and resource-adequacy studies size the reserve margin from the aggregate variability many such loads present. A load model that smooths over facility-scale variance therefore understates both the PAR a fleet of AI data centers presents and the reserve margin planners must carry to serve it reliably.

What makes an AI training facility difficult to model is that its power is structured, not smooth. A single training job runs on hundreds to thousands of accelerators (graphics processing units, GPUs) at once, and every accelerator must finish its share of the current step before any can start the next, because the step completes only once their results are combined---a cycle known as bulk-synchronous-parallel (BSP) execution. All accelerators perform dense forward and backward computation at thermal design power, halt together at a synchronization barrier to combine results, then periodically pause to checkpoint to storage~\cite{choukse2025power,go2025characterizing}. Each cycle produces a power swing between near-peak and near-idle at a timescale of seconds, and because the barrier is global, thousands of accelerators swing together instead of averaging each other out. This within-job structure is the source of the fast power transients that threaten frequency response and power quality.

A second mechanism shapes facility power one level above the individual job. A facility does not run a single job continuously, but instead runs a changing population of jobs that arrive, occupy blocks of nodes for hours to days, and depart~\cite{vercellino2026genai,majumder2026workloadCompositionAiDataCenters}. The number of occupied nodes rises through the working day, falls overnight, and varies across the week and the season, so the facility load carries a slow, large-amplitude modulation on top of the fast BSP transients. This slower, scheduling-driven modulation sets the facility-scale variance and the PAR that determine interconnection capacity and resource-adequacy risk.

Probabilistic load models for AI data centers have addressed the two mechanisms separately, and each strand leaves a gap. A whole-facility measurement and simulation approach~\cite{vercellino2026genai} supplies a discrete-event job scheduler and reproduces PAR above two at low utilization, but represents each job as a black-box power trace, so it cannot expose the within-job phase structure grid-stability studies require. A pattern-consistent dynamic load model~\cite{lu2026dynamic} similarly validates against facility measurements but uses a continuous utilization surrogate that omits BSP structure. On the within-job side, a wide-area oscillation study~\cite{ko2026widearea} models a training job as a two-phase semi-Markov chain with the number of concurrent jobs held fixed, so it carries no scheduling-driven facility variance. Closer to the scheduling side, a workload-composition study~\cite{majumder2026workloadCompositionAiDataCenters} models the scheduler in detail but reduces within-job power to statistical time-series templates fitted to past values, in place of a state machine.
Three gaps follow from this review. First, none of these models connects a multi-state within-job process to a job-arrival and scheduling process, so facility-scale variance has no physical generating mechanism and collapses when a node-level model is aggregated. Second, utilization is commonly treated as a fixed busy-node fraction and not as a fluctuating occupancy, which removes the dominant variance source by construction. Third, the within-job power model and the facility power model are seldom consistent representations of the same physics, so a model calibrated at one scale fails at the other.

This paper closes these gaps with three contributions. \emph{First}, it develops HSM-DC, a hierarchical semi-Markov load model that couples a within-job layer, consisting of a five-state semi-Markov chain over the phases of a BSP iteration and checkpoint writes and of barrier synchrony across the nodes of a job, to a job-scheduling between-job layer, so that node dynamics and IT power at both second and minute resolution derive from one consistent mechanism. \emph{Second}, it parameterizes the chain in the direction the physics fixes. The peak, base, and synchronization levels are anchored to the node thermal design power, the busy-node mean of the straight-line facility-power-versus-utilization relationship of a whole-facility reference, and the idle floor, and the sojourn times follow in closed form from the time budget and an iteration period placed in the band of measured multi-node racks, with no free fitting of the power levels. \emph{Third}, it validates the model at three scales: at facility scale against the reference across four utilization levels on the mean, standard deviation, PAR, autocorrelation, and queued-job fraction; at node scale against measured training-node telemetry; and at second scale against the band-integrated spectrum of a measured multi-node rack, together with a numerical check of the inter-facility covariance that the model's sharded-job mechanism predicts. The load model is presented in Section~2, validated in Section~3, and summarized in Section~4.

    \section{The HSM-DC Load Model}
    This section develops the HSM-DC load model, which decomposes facility power into IT and non-IT components and couples a job-scheduling layer with a within-job semi-Markov layer to generate the IT load. An AI data center is not a homogeneous load. It is a collection of physically distinct subsystems, each with its own power factor and characteristic timescale~\cite{Chaudhary2025DataCenterStability,Chaudhary2026Modal,choukse2025power}. The active power (in megawatts) at the grid-connection bus of facility $i$ decomposes into the IT load and a non-IT base load,
\begin{equation}
P_i(t) = P_{\mathrm{IT},i}(t) + P_{\mathrm{cool},i}(t) + P_{\mathrm{aux},i},
\label{eq:decomp}
\end{equation}
where $P_{\mathrm{IT},i}$ is the GPU and central processing unit (CPU) compute load, $P_{\mathrm{cool},i}$ is the cooling load of the air-handling units and the chiller plant, and $P_{\mathrm{aux},i}$ is the quasi-static auxiliary load of uninterruptible power supplies, lighting, networking, and distribution losses, all in megawatts. Table~\ref{tab:components} lists the component fractions, displacement factors, and timescales, which is why the decomposition matters. The cooling plant, buffered by thermal mass, responds far more slowly than the IT load, and each component draws reactive power at its own displacement angle. HSM-DC models the IT load through the two-layer mechanism below and represents the slower non-IT load through the cooling model at the end of this section.

\begin{table}[htb]
\centering
\caption{Facility load components: power fraction, displacement factor, and timescale \cite{Chaudhary2025DataCenterStability, Chaudhary2026Modal,choukse2025power}.}
\label{tab:components}
\small
\begin{tabular}{@{}llll@{}}
\toprule
\textbf{Component} & \textbf{Fraction} & \textbf{PF} & \textbf{Timescale} \\
\midrule
IT (GPU/CPU)        & 60\% & 0.995 at rated load & 0.1--30~s (BSP + scheduling) \\
HVAC air-handling   & 15\% & 0.940 & $\tau \approx 60$~s \\
Chiller plant       & 20\% & 0.950 & $\tau \approx 120$~s \\
Auxiliary           & 5\%  & 0.970 & quasi-static \\
\bottomrule
\end{tabular}
\end{table}

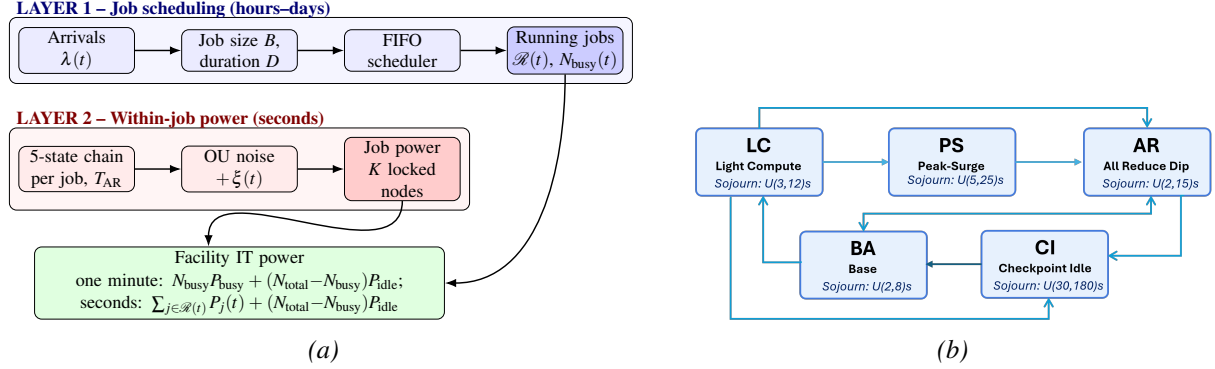
\begin{figure}[htb]
\centering
\begin{subfigure}[b]{0.52\linewidth}
\centering
\resizebox{\linewidth}{!}{%
\begin{tikzpicture}[
  font=\footnotesize,
  box/.style={draw, rounded corners, align=center, text width=21mm,
              minimum height=8mm, inner sep=2pt, fill=white},
  arr/.style={-{Latex[length=2.2mm]}, thick},
  title/.style={font=\footnotesize\bfseries}
]
\node[box, fill=blue!8] (a1) {Arrivals\\$\lambda(t)$};
\node[box, fill=blue!8, right=9mm of a1] (a2) {Job size $B$,\\duration $D$};
\node[box, fill=blue!8, right=9mm of a2] (a3) {FIFO\\scheduler};
\node[box, fill=blue!20, right=9mm of a3] (a4) {Running jobs\\$\mathcal{R}(t)$, $\Nbusy(t)$};
\draw[arr] (a1)--(a2); \draw[arr] (a2)--(a3); \draw[arr] (a3)--(a4);

\node[box, fill=red!8, below=14mm of a1] (b1) {5-state chain\\per job, $T_{\mathrm{AR}}$};
\node[box, fill=red!8, right=9mm of b1] (b2) {OU noise\\$+\,\xi(t)$};
\node[box, fill=red!20, right=9mm of b2] (b3) {Job power\\$K$ locked nodes};
\draw[arr] (b1)--(b2); \draw[arr] (b2)--(b3);

\begin{scope}[on background layer]
  \node[draw, rounded corners, fill=blue!4, inner sep=5pt, fit=(a1)(a4)] (L1) {};
  \node[draw, rounded corners, fill=red!4,  inner sep=5pt, fit=(b1)(b3)] (L2) {};
\end{scope}
\node[title, blue!45!black, anchor=west] at ([yshift=1.5mm]L1.north west)
  {LAYER 1 -- Job scheduling (hours--days)};
\node[title, red!50!black, anchor=west] at ([yshift=1.5mm]L2.north west)
  {LAYER 2 -- Within-job power (seconds)};

\node[box, fill=green!12, below=7mm of L2.south, text width=78mm, align=center] (fac)
  {Facility IT power\\[1pt]
   one minute: $\Nbusy P_{\mathrm{busy}}+(N_{\mathrm{total}}{-}\Nbusy)P_{\mathrm{idle}}$;\;
   seconds: $\sum_{j\in\mathcal{R}(t)}P_j(t)+(N_{\mathrm{total}}{-}\Nbusy)P_{\mathrm{idle}}$};

\draw[arr] (a4.south) to[out=-90,in=0] (fac.east);
\draw[arr] (b3.south) to[out=-90,in=90] ([xshift=-6mm]fac.north);
\end{tikzpicture}
}
\caption{}
\label{fig:architecture-a}
\end{subfigure}
\hfill
\begin{subfigure}[b]{0.44\linewidth}
\centering
\includegraphics[width=\linewidth]{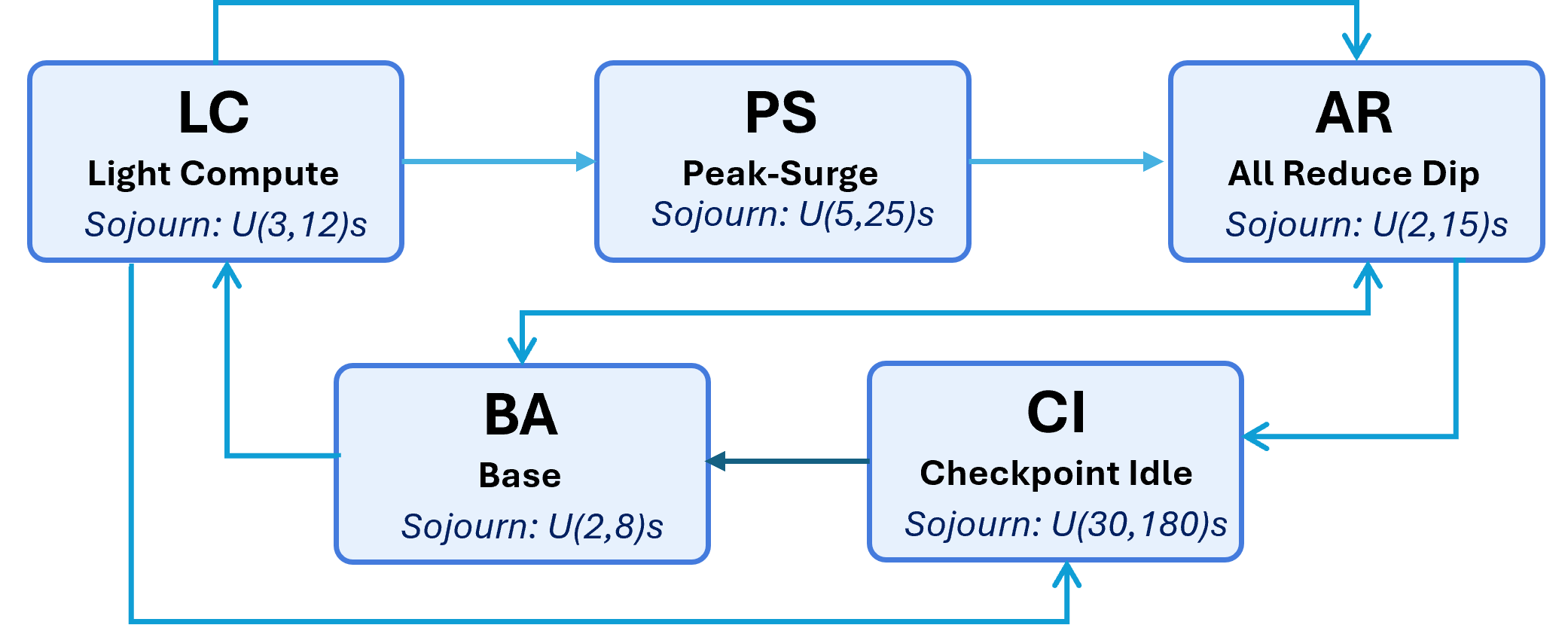}
\caption{}
\label{fig:architecture-b}
\end{subfigure}
\caption{HSM-DC architecture and within-job transitions. (a) Layer~1 (top) sets the running-job set $\mathcal{R}(t)$ and the busy-node count $\Nbusy(t)$; Layer~2 (bottom) drives one chain per job; the two combine into the IT power at minute and at second resolution. (b) Admissible transitions of the five-state semi-Markov chain (sojourn ranges shown are the submitted-model illustration; Table~\ref{tab:states} gives the re-solved sojourn means).}
\label{fig:architecture}
\end{figure}

The IT load is modeled by HSM-DC, a hierarchical model with two layers coupled through node occupancy. The job-scheduling layer (Layer~1) determines how many nodes are busy at each minute, and the within-job layer (Layer~2) determines how much power each busy node draws. The two layers separate by timescale. Layer~1 evolves over hours to days, set by job durations and the diurnal submission cycle, and Layer~2 evolves over seconds, set by the BSP iteration. At the minute-to-hour facility resolution used for adequacy studies the within-job chain averages to a busy-node mean and facility variance is carried by Layer~1. Below one minute the chain enters the IT power explicitly, one realization per running job, and supplies the BSP transients required for dynamic studies. Fig.~\ref{fig:architecture} shows the architecture.

\subsection{Job-Scheduling Layer}
Job submissions are non-stationary, rising after the start of the working day, peaking in the afternoon, and falling overnight~\cite{vercellino2026genai}. The arrival-event intensity $\lambda(t)$, in arrival events per minute, is the product of four unit-mean shaping factors about a base rate $\lambda_0$, also in arrival events per minute,
\begin{equation}
\lambda(t) = \lambda_0\,\phi_{\mathrm{hod}}(t)\,\phi_{\mathrm{dow}}(t)\,\phi_{\mathrm{moy}}(t)\,\phi_{\mathrm{seas}}(t),
\label{eq:intensity}
\end{equation}
where $\phi_{\mathrm{hod}}$ is the hour-of-day multiplier with a deep overnight trough and a broad afternoon plateau, $\phi_{\mathrm{dow}}$ is the day-of-week multiplier with weekday and weekend-evening emphasis, $\phi_{\mathrm{moy}}$ is the month-of-year campaign multiplier, and $\phi_{\mathrm{seas}}$ is a persistent annual sinusoid. The number of arrival events in minute $t$ is Poisson with mean $\lambda(t)$, and each event submits a batch of jobs, so the process is compound Poisson, which raises the variance of concurrent occupancy above the plain-Poisson level.

Each job carries a node count $B$ drawn from a heavy-tailed distribution on the power-of-two support $\{1,\ldots,256\}$, consistent with the parallel-workload characterization of rigid-job sizes~\cite{lublin2003parallel}. Most jobs therefore claim only a few nodes, but unusually large jobs occur far more often than a bell-shaped distribution would predict. Job duration $D$, in hours, is drawn from a two-component lognormal mixture~\cite{feitelson2015workload},
\begin{equation}
D \sim
\begin{cases}
\mathrm{Lognormal}(\mu_{\mathrm{short}}, \sigma_{\mathrm{short}}), & \text{w.p. } 1-f_{\mathrm{long}}, \\[3pt]
\mathrm{Lognormal}(\mu_{\mathrm{long}}, \sigma_{\mathrm{long}}), & \text{w.p. } f_{\mathrm{long}},
\end{cases}
\label{eq:duration}
\end{equation}
clipped to a finite support. The short component, whose typical duration is near one hour, represents fine-tuning, evaluation, and image-generation jobs that track the diurnal arrival swing and set the facility variance. The long component, a fraction $f_{\mathrm{long}}$ of all jobs with typical duration near twelve hours, represents multi-day pre-training campaigns that hold nodes for long horizons and create the persistent low-frequency baseline responsible for the high autocorrelation of the measured occupancy. The mixture resolves a tension that neither a single short nor a single long distribution satisfies. Short jobs alone produce variance without persistence, and long jobs alone produce persistence without variance.

Jobs are placed on a pool of $N_{\mathrm{total}}$ nodes by a first-in-first-out (FIFO) scheduler~\cite{vercellino2026genai}. A job starts at the first minute at or after its arrival at which the pool holds at least $B$ free nodes, holds its nodes for its duration, and then releases them. A head-of-line job that cannot fit blocks later jobs. The occupied-node count is
\begin{equation}
\Nbusy(t) = \sum_{j \in \mathcal{R}(t)} B_j,
\label{eq:Nbusy}
\end{equation}
where $\mathcal{R}(t)$ is the set of jobs running at minute $t$. Its variance decomposes by the random-sum identity,
\begin{equation}
\mathrm{Var}[\Nbusy] = \mathbb{E}[J]\,\mathrm{Var}[B] + \mathrm{Var}[J]\,\mathbb{E}[B]^2,
\label{eq:wald}
\end{equation}
where $J$ is the number of concurrent jobs. The identity assumes that the node counts $B_j$ of the jobs running at a given minute are independent and identically distributed and independent of $J$, an assumption that matches how HSM-DC draws job sizes. Each job's node count is sampled from the fixed size distribution introduced above without conditioning on its arrival time, its duration, or the number of other jobs running concurrently. The second term, the fluctuating concurrency, dominates and is the principal source of facility-power variance. The base rate $\lambda_0$ is not specified directly. It is calibrated by bisection to a target time-averaged node-occupancy fraction $\mu_{\mathrm{avg}} = (1/N_{\mathrm{total}})(1/T)\sum_t \Nbusy(t)$~\cite{vercellino2026genai}. Utilization in this definition is node occupancy, not compute or power utilization; because busy nodes rarely sit at thermal design power continuously (Section~2.2), full occupancy still falls well short of nameplate power, as Section~3's reference case quantifies.

\subsection{Within-Job Semi-Markov Layer}

Each running job drives its nodes through a semi-Markov process on the state space $\Scal=\{\PS,\LC,\BA,\CI,\AR\}$, the five power regimes of one BSP iteration:
\begin{enumerate}[leftmargin=1.4em,itemsep=1pt,topsep=2pt]
\item \textbf{Peak-Surge (\PS).} Dense forward and backward computation at thermal design power.
\item \textbf{Light-Compute (\LC).} The wait for other nodes or the accumulation of partial results, at power above the reference level but below \PS.
\item \textbf{Base (\BA).} The housekeeping gap between iterations, at the reference level $P_{\mathrm{busy}}$.
\item \textbf{Checkpoint-Idle (\CI).} The interval, thirty seconds to three minutes long, during which model weights and optimizer state are written to persistent storage while the accelerators idle.
\item \textbf{AllReduce-Dip (\AR).} The barrier at which every node of the job halts computation together to exchange gradients, the sharpest and lowest dip and the principal source of the grid-facing transient.
\end{enumerate}
Checkpointing is entered only from \AR, on a small branch probability (Fig.~\ref{fig:architecture}\subref{fig:architecture-b}), so it recurs roughly every ten to twenty minutes even though it lasts three orders of magnitude longer than an AllReduce. The two timescales coexist in one chain because the sojourn-time identity below ties \CI's mean duration to its own time budget, not to the fast states' sojourns.

The chain is parameterized in the direction that the physics fixes. The power levels are the known quantities: the \PS{} level is the node thermal design power $P^{\mathrm{TDP}}$, the \AR{} level is the measured idle floor $P_{\mathrm{idle}}$, the \BA{} level is the busy-node reference $P_{\mathrm{busy}}$ obtained from the whole-facility affine fit (Table~\ref{tab:anchors}), and the \CI{} level is fixed at $c_{\CI}=0.40$, within the checkpoint-idle range reported for accelerator checkpoint writes. Writing the level of state $s$ as $c_sP_{\mathrm{busy}}$ and the long-run fraction of time in state $s$ as $\pi^{\mathrm{sm}}_s$, the remaining level $c_{\LC}$ follows from the requirement that the time average of the node power equal $P_{\mathrm{busy}}$,
\begin{equation}
\sum_{s\in\Scal}\pi^{\mathrm{sm}}_s c_s = 1,
\label{eq:meananchor}
\end{equation}
where $\pi^{\mathrm{sm}}_{\CI}=0.075$ is the checkpoint duty cycle implied by a lognormal inter-checkpoint period (median 20~min, geometric standard deviation 1.5) and a uniform duration on 30--180~s. Because a node can fall about twice as far below its reference as it can rise above it ($1-c_{\AR}=0.83$ against $c_{\PS}-1=0.41$), \eqref{eq:meananchor} has an admissible solution $1\le c_{\LC}\le c_{\PS}$ only if $(c_{\PS}-1)\pi^{\mathrm{sm}}_{\PS}\le(1-c_{\AR})\pi^{\mathrm{sm}}_{\AR}+\varepsilon\le(c_{\PS}-1)(\pi^{\mathrm{sm}}_{\PS}+\pi^{\mathrm{sm}}_{\LC})$ with $\varepsilon=\pi^{\mathrm{sm}}_{\CI}(1-c_{\CI})$: time at the ceiling must be paid for by time near the floor, and time at the floor must not exceed what the high states can offset, net of the drag that \CI{} itself puts on the mean. The budget of Table~\ref{tab:states} satisfies both inequalities with margin.

\begin{figure}[htb]
\centering
\includegraphics[width=0.72\linewidth]{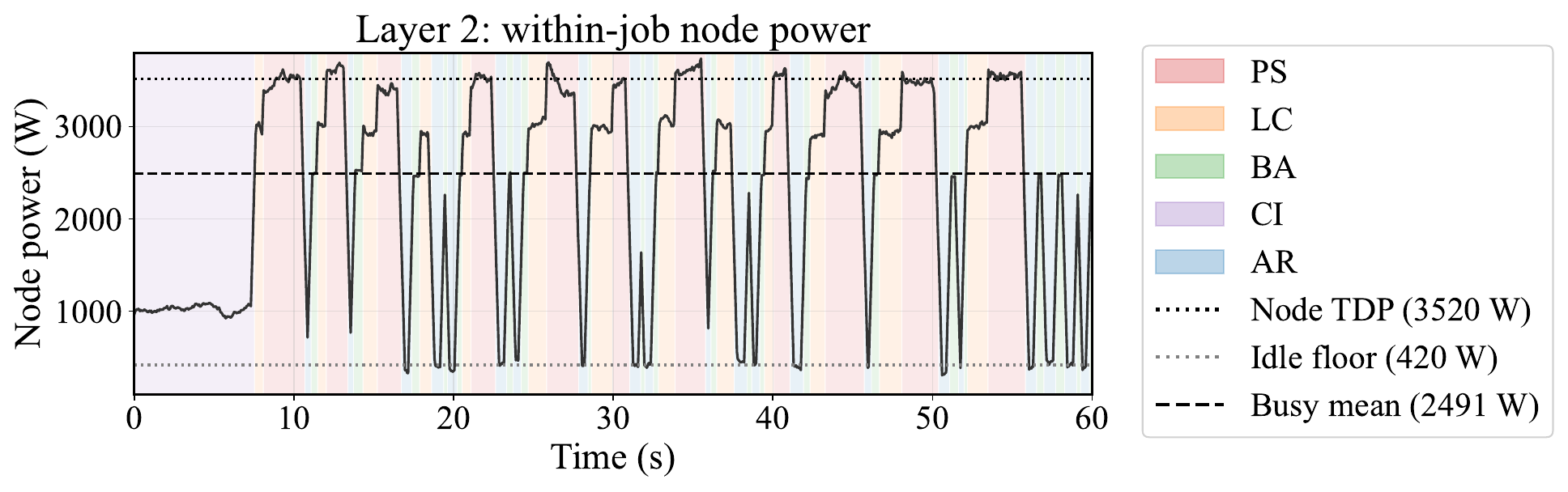}
\vspace{-0.8em}
\caption{Layer~2 within-job node power over 60~s at the median iteration period. The chain swings between thermal design power (\PS) and the idle floor (\AR) about three times per ten seconds, drops to the checkpoint level (\CI) for one write near $t=12$~s in this realization, and time-averages to the busy-node mean.}
\label{fig:nodetrace}
\end{figure}

The sojourn means follow from the Markov-renewal identity. With $\boldsymbol{\pi}^{\mathrm{emb}}$ the stationary vector of the embedded chain $\mathbf{P}$ (Fig.~\ref{fig:architecture}\subref{fig:architecture-b}), obtained by solving the balance equations with the normalization constraint as a linear system because the near-periodic BSP cycle makes an eigenvalue solution ill-conditioned, $\pi^{\mathrm{sm}}_s=\pi^{\mathrm{emb}}_s m_s/\sum_{s'}\pi^{\mathrm{emb}}_{s'}m_{s'}$, and the mean return time to \AR{} is the iteration period $T_{\mathrm{AR}}=\sum_{s}\pi^{\mathrm{emb}}_{s}m_{s}/\pi^{\mathrm{emb}}_{\AR}$, so that
\begin{equation}
m_s = T_{\mathrm{AR}}\,\pi^{\mathrm{emb}}_{\AR}\,\frac{\pi^{\mathrm{sm}}_s}{\pi^{\mathrm{emb}}_s},\qquad s\in\Scal,
\label{eq:sojourn}
\end{equation}
which fixes every sojourn mean from the time budget and the iteration period, with no search. Sojourns are uniform on $[0.5\,m_s,\,1.5\,m_s]$, a finite support chosen because an AllReduce cannot complete faster than the interconnect latency. The iteration period is drawn per job from a lognormal distribution with median 3~s and geometric standard deviation 1.5, truncated to 1--7.5~s so that the cycle rate $1/T_{\mathrm{AR}}$ lies in 0.13--1~Hz, the band in which a measured multi-node rack carries most of its variance (Section~3.3)~\cite{choukse2025power}; the duty cycle $\pi^{\mathrm{sm}}_{\AR}=0.185$ then gives a barrier duration of 0.555~s at the median period, broadly consistent with the 0.65~s barriers segmented from that trace. The node power in state $s$ is
\begin{equation}
P_{\mathrm{node}}(t) = \min\!\bigl[c_{s}P_{\mathrm{busy}} + \xi(t),\;P^{\mathrm{TDP}}\bigr],\qquad
d\xi = -\theta_{\mathrm{IT}}\,\xi\,dt + \sigma_{\mathrm{IT}}(s)P_{\mathrm{busy}}\,dW_t,
\label{eq:pnode}
\end{equation}
where $\xi(t)$ is an Ornstein--Uhlenbeck residual with mean-reversion rate $\theta_{\mathrm{IT}}=0.8~\mathrm{s}^{-1}$, integrated by its exact discrete-time solution~\cite{bibbona2008ornstein}, and the minimum represents the node power limiter. Level changes are linear ramps of 0.5~s centered on the transition instant, so that the trace is continuous and the time average of \eqref{eq:pnode} is unchanged. Table~\ref{tab:states} lists the resulting parameterization and Fig.~\ref{fig:nodetrace} shows a simulated node trace.

\begin{table}[htb]
\centering
\caption{Within-job state parameterization, in symbols that hold for any node and facility. Levels follow from~\eqref{eq:meananchor}; sojourn ranges follow from~\eqref{eq:sojourn}, and differ by state through $m_s$. $\sigma_{\mathrm{IT}}(s)$ is a per-state constant calibrated to the reference case (Table~\ref{tab:anchors}), which also gives numeric values for every quantity in this table.}
\label{tab:states}
\footnotesize
\renewcommand{\arraystretch}{1.1}
\setlength{\tabcolsep}{2.5pt}
\begin{tabular}{@{}l l p{3.0cm} c c c c@{}}
\toprule
\textbf{State} & \textbf{Full Name} & \textbf{BSP Phase} & \textbf{IT Power} & \textbf{Sojourn range} & \textbf{sgn} & $\sigma_{\mathrm{IT}}$ \\
\midrule
\PS & Peak-Surge      & Forward/backward propagation      & $P^{\mathrm{TDP}}$      & $[0.5,1.5]\,m_{\PS}$ & $+1$ & calibrated \\
\LC & Light-Compute   & Pipeline bubbles, gradient accumulation & solved from~\eqref{eq:meananchor} & $[0.5,1.5]\,m_{\LC}$ & $+1$ & calibrated \\
\BA & Base            & Inter-iteration gap                & $P_{\mathrm{busy}}$ (reference)      & $[0.5,1.5]\,m_{\BA}$ & $0$  & calibrated \\
\CI & Checkpoint-Idle & Checkpoint write, accelerators idle & $c_{\CI}P_{\mathrm{busy}}$  & $[0.5,1.5]\,m_{\CI}$ & $-1$ & calibrated \\
\AR & AllReduce-Dip   & Cross-node gradient synchronization & $P_{\mathrm{idle}}$     & $[0.5,1.5]\,m_{\AR}$ & $-1$ & calibrated \\
\midrule
\multicolumn{7}{l}{\footnotesize \PS{} and \AR{} are anchored to hardware; \BA{} is the reference level by definition;} \\
\multicolumn{7}{l}{\footnotesize \CI{} is fixed within the checkpoint-power range (Choukse et al.); \LC{} follows from~\eqref{eq:meananchor}.} \\
\multicolumn{7}{l}{\footnotesize sgn denotes the level's direction relative to \BA. $\CI$ is entered only from \AR{}, at} \\
\multicolumn{7}{l}{\footnotesize branch probability 0.005.} \\
\bottomrule
\end{tabular}
\setlength{\tabcolsep}{6pt}
\end{table}

\subsection{Coupling, Synchrony, and Calibration}

At minute resolution busy nodes draw the busy-node mean and unallocated nodes draw the idle power $P_{\mathrm{idle}}$ (in watts, on the same per-node basis as $P_{\mathrm{busy}}$),
\begin{equation}
P_{\mathrm{IT}}(t) = N_{\mathrm{total}}\,P_{\mathrm{idle}} + \Nbusy(t)\,\bigl(P_{\mathrm{busy}}-P_{\mathrm{idle}}\bigr),
\label{eq:facility}
\end{equation}
clipped at a saturation ceiling $P_{\mathrm{sat}}=\rho_{\mathrm{sat}}P^{\mathrm{rated}}$, where $P^{\mathrm{rated}}$ is the facility nameplate power and $\rho_{\mathrm{sat}}$ is a dimensionless fraction; $P_{\mathrm{IT}}$, $P_{\mathrm{sat}}$, and $P^{\mathrm{rated}}$ are reported in megawatts once the per-node terms are summed over $N_{\mathrm{total}}$ nodes. The mean IT power is therefore affine in occupancy, with intercept $N_{\mathrm{total}}P_{\mathrm{idle}}$ and slope $P_{\mathrm{busy}}-P_{\mathrm{idle}}$ that matches the reference's mean-versus-utilization relationship.

Below one minute the busy term is replaced by an explicit sum over the running jobs. The nodes of one job share the synchronization barrier, so their phases are locked, which is the mechanism that makes the AllReduce dip visible at the point of interconnection: independently phased nodes would average out by the central limit theorem. A synchrony fraction $S\in[0,1]$ sets the number of phase-locked nodes of job $j$ as the integer $K_j=\lfloor SB_j\rceil$; these nodes follow one realization of the chain, and the remaining $B_j-K_j$ nodes follow independent realizations. The second-scale IT power is
\begin{equation}
P_{\mathrm{IT}}(t) = \sum_{j\in\mathcal{R}(t)}\Bigl[K_j\,P^{(j)}_{\mathrm{node}}(t) + \sum_{n=1}^{B_j-K_j}P^{(j,n)}_{\mathrm{node}}(t)\Bigr]
+ \bigl(N_{\mathrm{total}}-\Nbusy(t)\bigr)P_{\mathrm{idle}},
\label{eq:facility_sec}
\end{equation}
with every locked node drawing the checkpoint level whenever the shared chain is in state \CI, and without a saturation clip, because the levels are anchored to hardware. For a fixed set of running jobs, the variance of the phase-and-residual component is $\sum_j[\sigma_{\mathrm{ph}}^2(K_j^2+B_j-K_j)+\sigma_{\mathrm{ou}}^2B_j]$, where $\sigma_{\mathrm{ph}}^2$ is the stationary variance of the phase level of one node, including its excursions to \CI, and $\sigma_{\mathrm{ou}}^2$ is the variance of its residual. The synchronized part therefore grows with $K_j^2$, so the second-scale standard deviation of the facility falls with the square root of the busy-node count. $S=1$ is full barrier synchrony and $S=0$ the independent-node limit. Time-averaging \eqref{eq:facility_sec} over a minute returns \eqref{eq:facility} because the chain averages to $P_{\mathrm{busy}}$ by \eqref{eq:meananchor}, which Section~3.3 verifies numerically.

The model above is anchored, not freely fit. Every level and rate is either a hardware fact or a measured reference statistic, with no free parameter tuned to match the validation results. Section~3 introduces the one reference case used throughout this paper and lists the anchors and sources (Table~\ref{tab:anchors}).

\subsection{Non-IT Components, Cooling Dynamics, and Reactive Power}

The cooling load is not a static fraction of the IT load. Instead, the cooling electrical power is a delayed, low-pass-filtered response to the IT load, not an instantaneous copy of it~\cite{bessasparis2025forecasting,nerc2025largeloads}. This separation of timescales is the defining feature of the non-IT load. It filters the second-scale BSP and scheduling transients out of the cooling power, so the facility bus carries fast IT swings, a smooth lagged cooling load, and a constant auxiliary load.

The cooling load is modeled as a first-order thermal-mass response of each cooling sub-system to the normalized IT load $u_{\mathrm{IT}}(t)=P_{\mathrm{IT}}(t)/P_{\mathrm{IT}}^{\mathrm{rated}}$, scaled by an ambient-temperature factor,
\begin{equation}
P_{\mathrm{cool}}(t) = \sum_{c\in\{\mathrm{HVAC},\mathrm{Ch}\}} \alpha_c\,\gamma\!\left(T_{\mathrm{amb}}(t)\right)\,\mathcal{L}_{\tau_c,\,\theta_d}\!\left[u_{\mathrm{IT}}\right]\!(t)\,P^{\mathrm{rated}},
\label{eq:cool}
\end{equation}
where $\alpha_{\mathrm{HVAC}}$ and $\alpha_{\mathrm{Ch}}$ are the nominal HVAC and chiller fractions (Table~\ref{tab:components}), $\mathcal{L}_{\tau_c,\theta_d}$ is a first-order lag of time constant $\tau_c$ and transport dead time $\theta_d$, both in seconds, and $\gamma$ is the dimensionless ambient-temperature multiplier. $P_{\mathrm{IT}}^{\mathrm{rated}}$ is the facility's IT nameplate power, $N_{\mathrm{total}}$ nodes at thermal design power. The HVAC time constant is shorter than the chiller time constant (Table~\ref{tab:components}), and a transport dead time $\theta_d=90$~s precedes the response. The temperature factor and the ambient temperature are~\cite{bessasparis2025forecasting,sun2024energy}
\begin{equation}
\gamma(T)=\max\!\left[0.1,\;1+\beta_T\frac{T-T_{\mathrm{ref}}}{\Delta T}\right],\quad
T_{\mathrm{amb}}(t)=T_{\mathrm{mean}}+\Delta T\cos\!\frac{2\pi(h-h_{\mathrm{summer}})}{8760}+T_{\mathrm{diu}}(t)+\theta_{\mathrm{c}}(t)+\theta_{i}(t),
\label{eq:Tamb}
\end{equation}
with dimensionless cooling sensitivity $\beta_T=0.25$, reference temperature $T_{\mathrm{ref}}=15~^\circ$C, seasonal amplitude $\Delta T=12~^\circ$C, annual mean $T_{\mathrm{mean}}=12~^\circ$C, summer peak $h_{\mathrm{summer}}=4680$~h, a diurnal term $T_{\mathrm{diu}}$ of half-amplitude 4~$^\circ$C peaking at 15:00, and two Ornstein--Uhlenbeck anomalies with two-hour time constants and standard deviation 1~$^\circ$C, of which $\theta_{\mathrm{c}}$ is shared by every facility of a region and $\theta_i$ is independent per facility. The lower bound 0.1 is the free-cooling floor. The auxiliary load is $P_{\mathrm{aux}}=\alpha_{\mathrm{Aux}}P^{\mathrm{rated}}$ with $\alpha_{\mathrm{Aux}}=0.05$. Each component draws reactive power at its displacement angle, $Q_c=P_c\tan\varphi_c$ with $\cos\varphi_c$ from Table~\ref{tab:components}, positive for lagging; because power-factor-corrected server supplies hold their factor near unity at rated load and their input filter contributes a leading component at light load, the IT angle is loading dependent, $\varphi_{\mathrm{IT}}(u)=\varphi_{\mathrm{IT}}^{\mathrm{rated}}-\kappa_\varphi(1-u)$ with $\kappa_\varphi=0.15$~rad, a stated assumption to be swept in grid studies.

\begin{figure}[htb]
\centering
\includegraphics[width=0.85\linewidth]{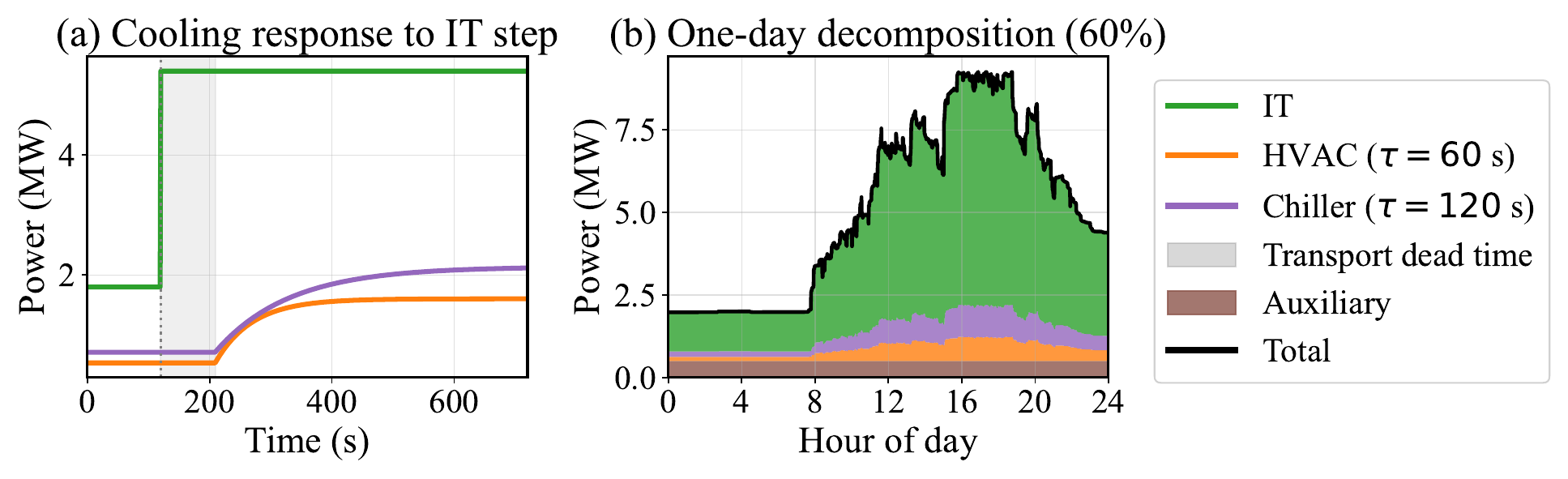}
\vspace{-1em}
\caption{Non-IT cooling dynamics. Left: response to an IT step, showing the transport dead time and the HVAC and chiller time constants. Right: one-day decomposition at 60\% utilization, where cooling lags the IT envelope while the IT load carries the fast variability.}
\label{fig:nonit}
\end{figure}

Fig.~\ref{fig:nonit} shows the step response and one-day decomposition. Because the reference whole-facility data is IT-only, the cooling layer is grounded in the component fractions, time constants, and temperature sensitivity reported in the data-center load literature~\cite{bessasparis2025forecasting,nerc2025largeloads,sun2024energy} and is not fitted to the reference; its validation against metered cooling power is identified as future work.

\subsection{Spatial Load Correlation Across Facilities}

AI data centers that share a workload-orchestration platform, a region, or a weather pattern do not draw power independently. Coordinated scheduling synchronizes barriers across sites, shared ambient conditions align cooling cycles, and common corridors couple converter dynamics; these interactions produce correlated load fluctuations that classical planning treats as independent~\cite{chaudhary2026pesgm}. This correlation is episodic, as sliding-window modal analysis of measured inter-bus coherence confirms~\cite{Chaudhary2026Modal}, and even moderate synchronization inflates aggregate variance beyond what diversity factors predict~\cite{chaudhary2026pmaps_adequacy}. Because IT power is generated from the job process, HSM-DC introduces this dependence at the job level, where its covariance is known, through three mechanisms whose zero setting recovers independent facilities that share only the deterministic diurnal profile. \emph{Sharded jobs}: a fraction $F_{\times}$ of jobs of 64 nodes or more is split across facilities of the same cluster; the shards share one start time and one chain realization, which also synchronizes their checkpoints, so that for a fixed job set the second-scale covariance of facilities $i$ and $l$ is $\sum_{j\ \mathrm{shared}}\sigma_{\mathrm{ph}}^2K_{j,i}K_{j,l}$, computable from the realized shard sizes. \emph{Cluster arrival bursts}: a common Poisson event stream per cluster is thinned to each member with probability $p_{\mathrm{burst}}$ and each member's own rate is reduced by the same expected amount, so campaigns start jobs at several facilities within the same minute while every facility's marginal arrival rate is unchanged. \emph{Shared weather}: the regional anomaly $\theta_{\mathrm{c}}$ of~\eqref{eq:Tamb}. A slow channel correlates the monthly multipliers $\phi_{\mathrm{moy}}$ through a Gaussian copula with a class-based correlation matrix, projected to the nearest correlation matrix when needed~\cite{higham2002nearestPSD}. This coupling carries a direct planning consequence: when concurrent training campaigns synchronize barrier and checkpoint events across facilities on adjacent substations, the aggregate peak departs upward from the value an independence-based diversity factor would predict, and the spinning reserve carried against the loss of the largest in-feed must be sized against the correlated, not the independent, aggregate~\cite{chaudhary2026pmaps_adequacy}.

    \section{Validation}
    \subsection{The Reference Case}

The 2840-node, 10~MW reference colocation facility~\cite{vercellino2026genai} is considered here; every number reported from this point onward therefore describes this specific case, not a universal constant. Table~\ref{tab:anchors} lists the anchors and sources; a different node or facility would carry these equations through to different values.

\begin{table}[htb]
\centering
\caption{Calibration anchors and sources, for the reference case evaluated in this section.}
\label{tab:anchors}
\small
\begin{tabular}{@{}lll@{}}
\toprule
\textbf{Quantity} & \textbf{Value} & \textbf{Source} \\
\midrule
Node thermal design power & 3520~W & 4$\times$H100 + 2$\times$EPYC~\cite{vercellino2026genai} \\
Idle floor per node & 420~W & GPU+CPU idle~\cite{vercellino2026genai} \\
Busy-node mean & 2491~W & reference affine slope + idle \\
Facility mean & $1.2035+5.883\,\mu_{\mathrm{avg}}$ MW & reference fit, $R^2{=}1.0$ \\
Saturation ceiling (one-minute path) & 73.3\% of nameplate & reference plateau \\
Iteration period band & 0.13--1~Hz & multi-node rack trace~\cite{choukse2025power} \\
\bottomrule
\end{tabular}
\end{table}

These anchors, carried through~\eqref{eq:meananchor}, are what fix Table~\ref{tab:states}'s per-unit levels to actual values for this case (e.g.\ $c_{\PS}=1.414$) and, through~\eqref{eq:sojourn}, its sojourn ranges (e.g.\ $\CI$: U(22.38, 67.13)~s) -- the resolved parameterization the rest of this section validates.

\subsection{Facility-Scale Validation}

Each configuration is simulated for one year at minute resolution at each of four target utilizations, and the output is compared to the reference at the same utilization. A single scalar cannot capture whether two year-long power records agree, because two records can share the same mean and still differ in their tails, their timing, or their minute-to-minute swings. A multi-axis battery of metrics is used instead: the 1-Wasserstein distance on the power marginal for level fidelity, PAR and high percentiles for the tail, the $L_2$ distance between autocorrelation functions for temporal structure, and the ramp distribution for short-horizon variability~\cite{gneiting2007scoring}.

Table~\ref{tab:valid} reports the comparison. The mean is reproduced to within 20~kW at every utilization, with a coefficient of determination of 0.9997 across utilization. The standard deviation is reproduced with $R^2=0.92$ and PAR with $R^2=0.82$. The model also reproduces the queued-job fraction, 76.0\% at 80\% utilization against the reference 75.6\%, and the monotonic decline of PAR with utilization. This decline is the operational signature of the facility. At low utilization, the mean is dominated by the idle floor while diurnal submission bursts still drive high peaks, but as utilization rises, the mean climbs toward the saturation ceiling and the ratio falls.

\begin{table}[htb]
\centering
\caption{HSM-DC versus the whole-facility reference at matched scale (2840 nodes, 10~MW).}
\label{tab:valid}
\small
\begin{tabular}{@{}lcccccc@{}}
\toprule
\textbf{Util.} & \multicolumn{2}{c}{Mean (MW)} & \multicolumn{2}{c}{Std (MW)} & \multicolumn{2}{c}{PAR} \\
 & data & model & data & model & data & model \\
\midrule
20\% & 2.379 & 2.367 & 0.848 & 0.844 & 2.558 & 2.981 \\
40\% & 3.557 & 3.563 & 1.591 & 1.457 & 2.063 & 1.987 \\
60\% & 4.736 & 4.702 & 1.907 & 1.799 & 1.544 & 1.506 \\
80\% & 5.908 & 5.882 & 1.776 & 1.627 & 1.239 & 1.204 \\
\midrule
\multicolumn{7}{l}{$R^2(\mathrm{mean})=0.9997$,\;\; $R^2(\mathrm{std})=0.92$,\;\; $R^2(\mathrm{PAR})=0.82$} \\
\bottomrule
\end{tabular}
\end{table}

\begin{figure}[htb]
\centering
\includegraphics[width=0.9\linewidth]{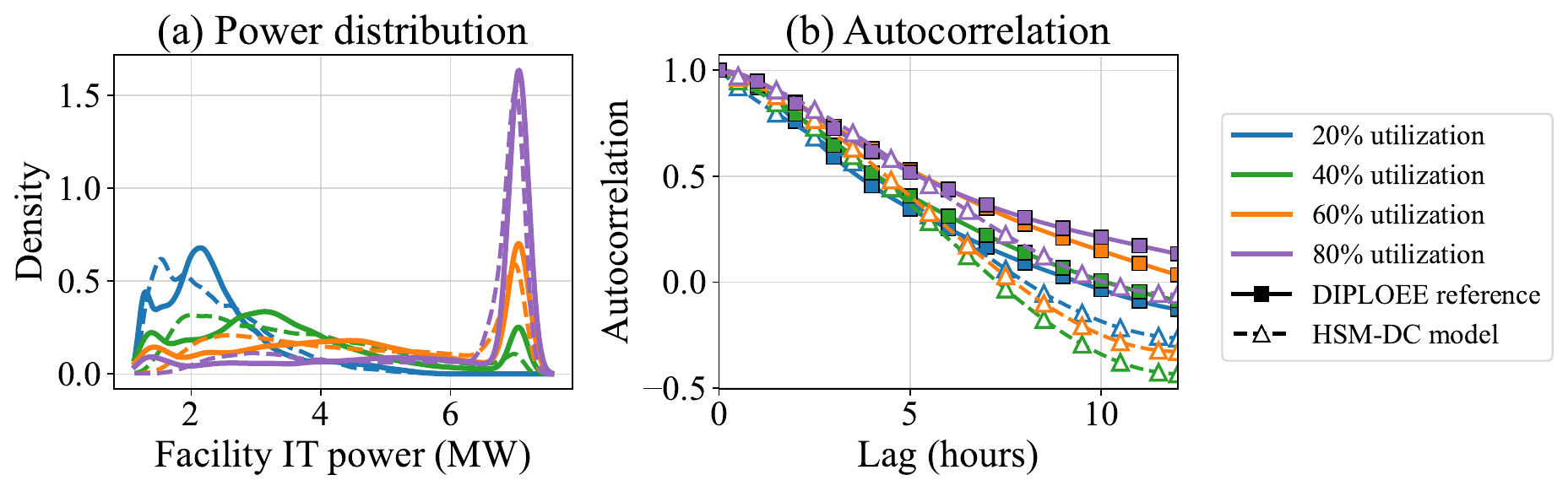}
\vspace{-1em}
\caption{IT power distribution (a) and autocorrelation (b) across four utilization levels, colored by utilization.}
\label{fig:dist_acf}
\end{figure}

\begin{figure}[htb]
\centering
\includegraphics[width=0.9\linewidth]{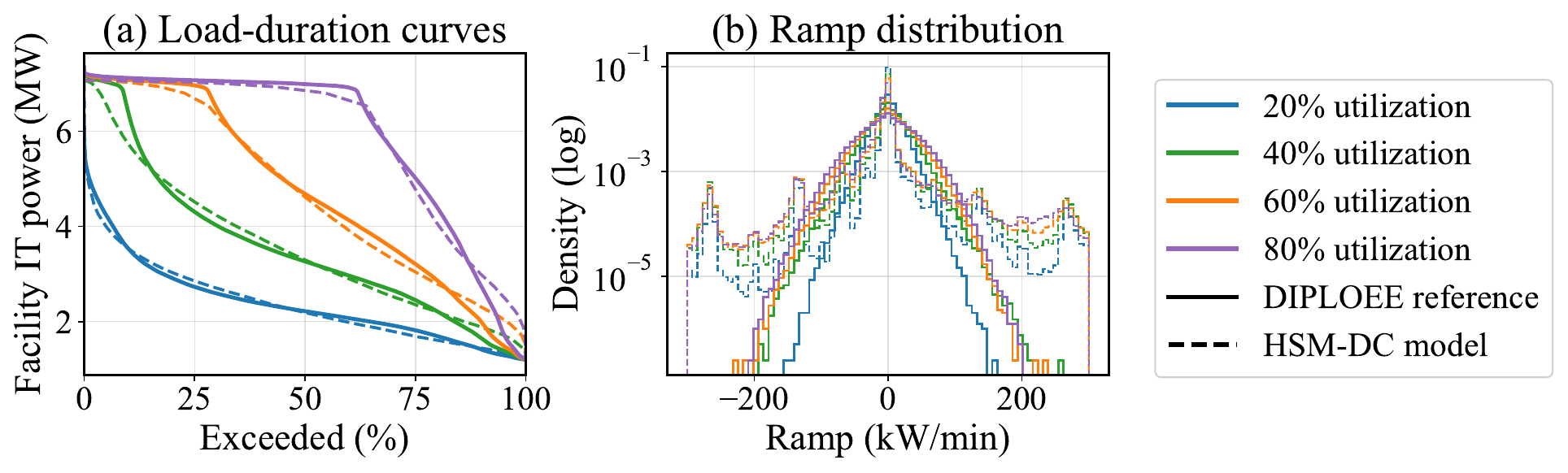}
\vspace{-1em}
\caption{Load-duration and ramp behavior across four utilization levels: (a) load-duration curves and (b) one-minute ramp distribution.}
\label{fig:ldc_ramp}
\end{figure}

The distributional and temporal agreement is shown side by side in Fig.~\ref{fig:dist_acf}. At each utilization the model reproduces the shape of the reference marginal, including the right-skewed body at low utilization and the saturation pile-up that grows as utilization rises, and the model autocorrelation tracks the reference over lags of hours and decays slowly because the long-job component holds occupancy for extended horizons. The load-duration and ramp behavior, which govern resource adequacy and reserve sizing, are reproduced in Fig.~\ref{fig:ldc_ramp}. The load-duration curves match across the full exceedance range at every utilization, including the high-load tail that drives loss-of-load risk, and the one-minute ramp distribution matches in both the bulk and the tail at every utilization.

\subsection{Independent Node-Level Validation}

Because the facility validation above depends on a simulated whole-facility reference, the within-job layer admits an independent check against direct measurement. Elsayed~\textit{et al.}~\cite{elsayed2026characterization} report 50~Hz power telemetry of eight-GPU H100 nodes during Llama-8B training, on hardware distinct from the four-GPU node used to calibrate HSM-DC. The comparison is made on a per-unit basis using the node thermal design power so that the eight-GPU and four-GPU nodes are commensurable. The measured node synchronizes its accelerators over the intra-node interconnect, so the comparison tests the level structure of the chain, the bimodal distribution, the compute plateau, and the depth of the synchronization dip, and not the iteration period, which the model sets from multi-node evidence.

\begin{figure}[htb]
\centering
\includegraphics[width=0.9\linewidth]{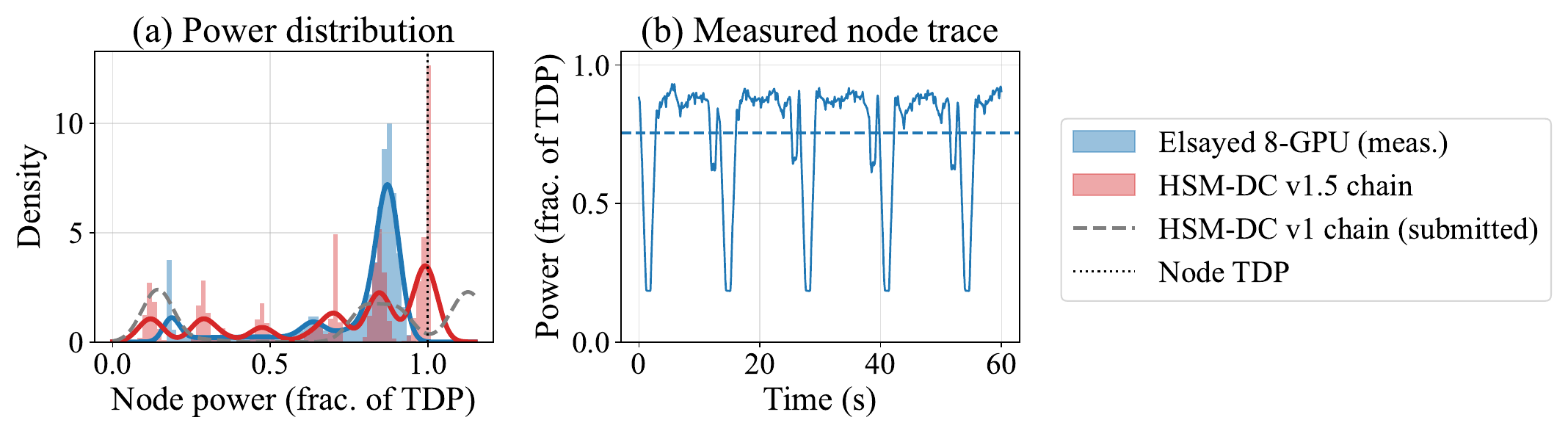}
\vspace{-1em}
\caption{Independent node-level validation. Left: bimodal node-power distribution, HSM-DC Layer~2 against the measured eight-GPU H100 node, in per-unit of node thermal design power. Right: a measured 60~s node trace showing the compute plateau and the synchronization dips that the chain encodes.}
\label{fig:node}
\end{figure}

Fig.~\ref{fig:node} shows the result. The measured trace exhibits the bulk-synchronous-parallel signature the chain encodes, a sustained compute plateau punctuated by sharp dips at the synchronization barriers. The model reproduces the bimodal distribution, with a compute mode at 84.8\% of thermal design power against the measured 82.5\%, a low mode at 21.7\% against the measured 24.9\%, and 22\% of time in the low mode against the measured 12\%; the model maximum is the thermal design power by construction of the power limiter in~\eqref{eq:pnode}. The overall distributional match is a 1-Wasserstein distance of 0.11 per unit. The remaining differences are physically interpretable and conservative: the model dips to the full idle floor while the measured node continues to draw power for background communication, and the model spends more time in the low mode because it targets multi-node pre-training, where barriers are more frequent than in the measured fine-tuning run. The node-level agreement, on independent hardware and a measured trace, corroborates the within-job layer that the facility validation cannot isolate.

\subsection{Second-Scale Spectral Validation}

\begin{figure}[htb]
\centering
\includegraphics[width=0.9\linewidth]{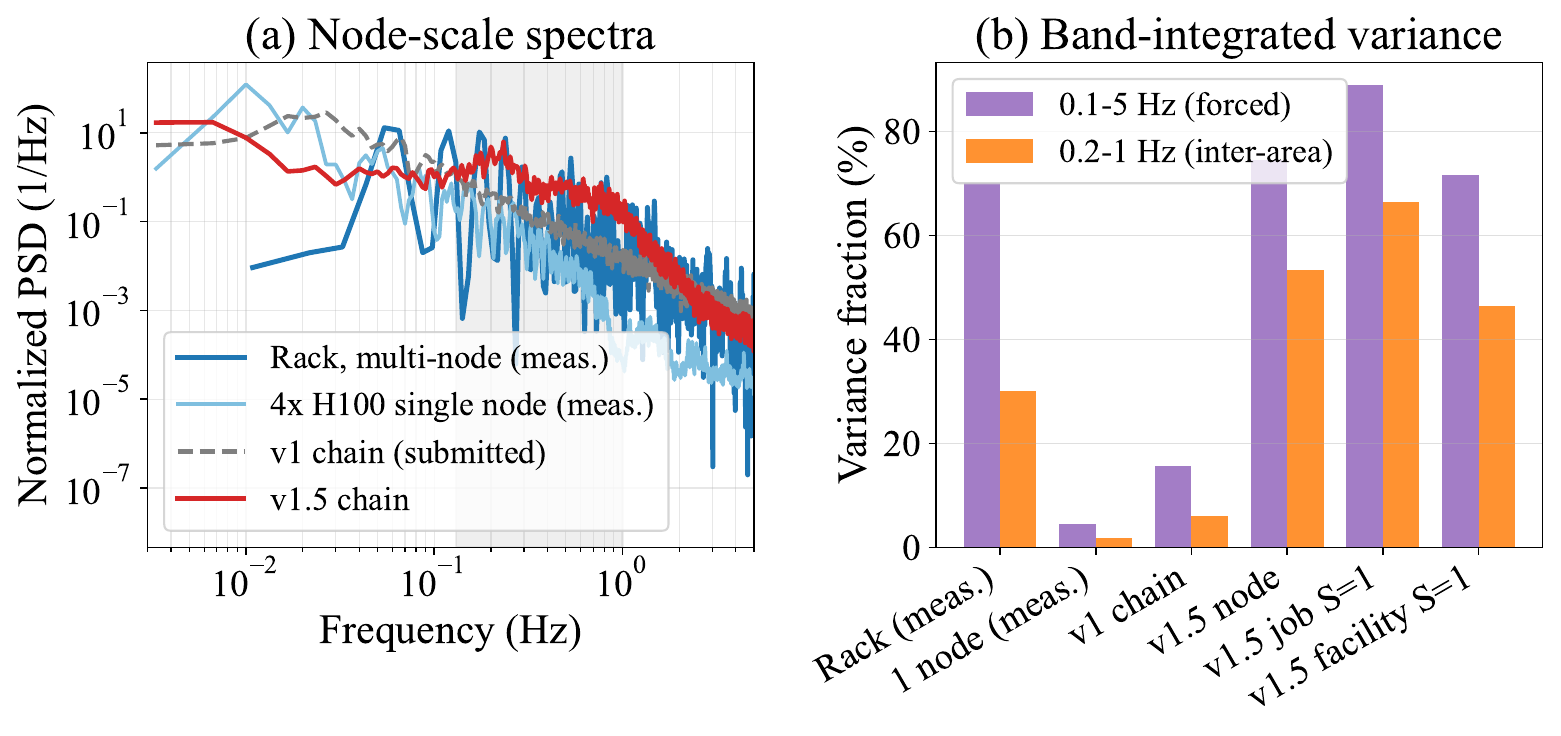}
\vspace{-1em}
\caption{Spectral placement of the within-job layer. (a) Normalized power spectra of the measured rack and single-node traces and of the HSM-DC chain with and without the frequency anchor; the shaded band is 0.13--1~Hz. (b) Band-integrated variance fractions.}
\label{fig:spectrum}
\end{figure}

The second-scale path of~\eqref{eq:facility_sec} is checked in three ways. First, the frequency band in which the generated power carries its variance is compared with measured GPU power traces by band-integrated Welch spectra, which report the fraction of the signal variance in each band without picking peaks. The measured references are six single-node fine-tuning traces at 10~Hz~\cite{yan2026switching} and one multi-node DGX-H100 rack trace digitized from the power-stabilization study of Choukse~\textit{et al.}~\cite{choukse2025power}, which is normalized and 152~s long and therefore supports the band placement, not the amplitude. Fig.~\ref{fig:spectrum} shows the result. The single-node measurements, which synchronize inside the node, carry under 4\% of their variance in 0.2--1~Hz, whereas the multi-node rack carries 70\% in 0.1--5~Hz and 30\% in 0.2--1~Hz. A chain parameterized by hardware plausibility alone, with sojourns of 2--25~s and a cycle fundamental of 0.036~Hz, carries 16\% and 6\% in those bands and sits with the single-node traces; the chain of Table~\ref{tab:states} places 74\% and 53\% of a node's variance there and, for the reference facility at 60\% occupancy with full synchrony, 72\% and 46\% of the second-scale facility variance. With independent nodes ($S=0$) the same facility carries 21\% and 12\%. The fast phase swings still average out, but each node's own, independently timed checkpoint dip does not.

Second, the two facility-power paths are mutually consistent. On the reference facility at 60\% occupancy, the minute averages of~\eqref{eq:facility_sec} reproduce the one-minute path~\eqref{eq:facility} to $+0.35\%$ in the mean; this confirms that the second-scale layer adds variability without shifting the calibration of Section~3.1. Third, the coupling truth of the sharded-job mechanism is verified numerically: the phase covariance of a four-facility fleet with jobs sharded across two clusters agrees with the formula of Section~2.5 to within 7\% (ratios 0.944 and 1.065), and is zero within its standard error for facilities that do not share jobs.

\subsection{Limitations}

Two quantitative gaps remain in the facility-scale fit. The standard deviation runs about 8--9\% below the reference at intermediate and high utilization because the model generates a marginally less bursty arrival process than the reference trace. PAR at 20\% utilization runs higher than the reference because the model predicts that a facility averaging 20\% occupancy approaches saturation on its most extreme day, whereas the reference arrival trace peaks at 83\% occupancy. Both discrepancies point the same way. The model is slightly conservative at low utilization and predicts higher peak demand, the safe direction for resource-adequacy and interconnection assessment.

The validation is also bounded in scope. The reference IT power comes from a discrete-event simulation, not from metered grid data, so the comparison establishes consistency with a measurement-grounded reference and not with a physical facility. Closing this gap requires active and reactive power telemetry at the point of interconnection of an operating AI data center, once available from a hosting utility. The only multi-node trace available for the band placement is digitized, normalized, and 152~s long, so the in-band amplitude at facility scale and the synchrony fraction $S$ are not measured anywhere public and are carried as parameters. The cooling layer and the displacement-angle model are grounded in established parameters and stated assumptions rather than validated against metered non-IT or reactive power. Because the cooling response is a unity-gain first-order lag, halving or doubling $\tau_{\mathrm{HVAC}}$ and $\tau_{\mathrm{Ch}}$ changes the validated one-minute mean IT power, standard deviation, and PAR by less than 0.1\%, since the lag shifts timing, not the steady-state statistics; the sub-minute transient shape it governs is topology-dependent and is not validated here.

    \section{Conclusion}
    This paper has developed HSM-DC, a hierarchical semi-Markov load model for AI data centers that couples a job-scheduling layer with a within-job layer consisting of a five-state semi-Markov chain over the phases of a BSP iteration and checkpoint writes and of barrier synchrony across the nodes of a job. Tests against a whole-facility reference, measured node telemetry, and a measured multi-node rack confirm that the job-arrival and scheduling process is the origin of facility-scale variance and peak demand, and that synchrony within jobs is the origin of the second-scale swings visible at the point of interconnection. Because HSM-DC is anchored to measured node hardware specifications and the aggregate statistics of a reference facility, without being fit to a single proprietary dataset, it can generate realistic, statistically validated load scenarios in place of telemetry most utilities cannot obtain, including active and reactive setpoints per facility with a known inter-facility covariance. This supports interconnection capacity studies for prospective campuses, resource-adequacy analysis under correlated multi-facility growth, and sub-second dynamic studies of frequency response and protection coordination. The active and reactive power series HSM-DC produces attach to a standard time-varying load or current-injection model at the point of interconnection in RMS platforms such as PSS/E and PowerFactory or EMT platforms such as PSCAD, so a planner can drive an existing network model directly from HSM-DC output.

Future work can close the gaps the validation section identifies against metered telemetry, measure the synchrony fraction and the in-band amplitude at facility scale, and extend the job queue to represent an inference-serving request process.

    \section*{Acknowledgment}
    This article has been authored by employees of National Technology \& Engineering Solutions of Sandia, LLC under Contract No. DE-NA0003525 with the U.S. Department of Energy (DOE). The employees own all right, title and interest in and to the article and is solely responsible for its contents. The United States Government retains and the publisher, by accepting the article for publication, acknowledges that the United States Government retains a non-exclusive, paid-up, irrevocable, world-wide license to publish or reproduce the published form of this article or allow others to do so, for United States Government purposes. The DOE will provide public access to these results of federally sponsored research in accordance with the DOE Public Access Plan
\href{https://www.energy.gov/downloads/doe-public-access-plan}{https://www.energy.gov/downloads/doe-public-access-plan}.

    \begingroup
    \footnotesize
    \let\oldthebibliography\thebibliography
    \renewcommand{\thebibliography}[1]{%
        \oldthebibliography{#1}%
        \setlength{\itemsep}{0pt}%
        \setlength{\parsep}{0pt}%
        \setlength{\parskip}{0pt}%
    }
    \bibliographystyle{ieeetr}
    \bibliography{references}

@article{vercellino2026genai,
  author  = {Vercellino, C. and Willard, S. and Campos, J. and {da Silva Pereira}, R. and Hull, A. and Selensky, E. and Mueller, F.},
  title   = {Measurement of Generative {AI} Workload Power Profiles for Whole-Facility Data Center Infrastructure Planning},
  journal = {arXiv preprint arXiv:2604.07345},
  year    = {2026}
}

@article{elsayed2026characterization,
  author  = {Elsayed, Ahmed Abd Elaziz and Al-Obaidi, Abdullah Azhar and Farag, Hany E. Z.},
  title   = {Characterization of high-resolution {AI} data center training workloads on single and multiple {GPU} nodes},
  journal = {Scientific Data},
  year    = {2026},
  type    = {Data Descriptor},
  doi     = {10.1038/s41597-026-07496-6},
  url     = {https://doi.org/10.1038/s41597-026-07496-6},
  note    = {3 June 2026}
}

@inproceedings{go2025characterizing,
  title     = {{Characterizing the Efficiency of Distributed Training: A Power, Performance, and Thermal Perspective}},
  author    = {Go, Seokjin and Park, Joongun and More, Spandan and Wu, Hanjiang and Wang, Irene and Jezghani, Aaron and Krishna, Tushar and Mahajan, Divya},
  booktitle = {Proc. 58th IEEE/ACM Int. Symp. Microarchitecture (MICRO)},
  pages     = {626--642},
  year      = {2025}
}

@article{choukse2025power,
  title   = {{Power Stabilization for AI Training Datacenters}},
  author  = {Choukse, Esha and Warrier, Brijesh and Heath, Scot and Belmont, Luz and Zhao, April and Khan, Hassan Ali and Harry, Brian and Kappel, Matthew and Hewett, Russell J. and Datta, Kushal and others},
  journal = {arXiv preprint arXiv:2508.14318},
  year    = {2025}
}

@misc{lu2026dynamic,
  author        = {Lu, Siyu and Xiao, Chenhan and Weng, Yang},
  title         = {Dynamic Load Model for Data Centers with Pattern-Consistent Calibration},
  year          = {2026},
  month         = feb,
  eprint        = {2602.07859},
  archivePrefix = {arXiv},
  primaryClass  = {cs.LG},
  doi           = {10.48550/arXiv.2602.07859},
  url           = {https://arxiv.org/abs/2602.07859},
  note          = {Submitted February 8, 2026}
}

@article{ko2026widearea,
  author  = {Ko, Min-Seung and Zhu, Hao},
  title   = {Wide-Area Power System Oscillations from Large-Scale {AI} Workloads},
  journal = {{IEEE} Transactions on Power Systems},
  year    = {2026},
  pages   = {1--14},
  doi     = {10.1109/TPWRS.2026.3685506}
}

@misc{majumder2026workloadCompositionAiDataCenters,
  author        = {Majumder, Subir and Yu, Minlan and Xie, Le},
  title         = {Workload Composition Smooths Aggregate Power Demand while Sustaining Short-Horizon Ramps in {AI} Data Centers},
  year          = {2026},
  month         = apr,
  eprint        = {2604.10769},
  archivePrefix = {arXiv},
  primaryClass  = {eess.SY},
  doi           = {10.48550/arXiv.2604.10769},
  url           = {https://arxiv.org/abs/2604.10769},
  note          = {Submitted April 12, 2026}
}

@article{sheng2026power,
  author  = {Sheng, Yu and Zhang, Chenxuan and Zhu, Zixuan and Xu, Hongyi and Wen, Junqi and Wang, Ruoheng and Yang, Jianjun and Wang, Qin and Bu, Siqi},
  title   = {Power for {AI} Data Centers: Energy Demand, Grid Impacts, Challenges and Perspectives},
  journal = {Energies},
  year    = {2026},
  volume  = {19},
  number  = {3},
  article-number = {722},
  doi     = {10.3390/en19030722},
  url     = {https://doi.org/10.3390/en19030722}
}

@techreport{shehabi2024lbnl,
  title        = "{2024 United States Data Center Energy Usage Report}",
  author       = {Shehabi, Arman and Newkirk, Alex and Smith, Sarah J. and Hubbard, Alex and Lei, Nuoa and Siddik, Md Abu Bakar and Holecek, Billie and Koomey, Jonathan and Masanet, Eric and Sartor, Dale},
  institution  = {Lawrence Berkeley National Laboratory},
  year         = {2024},
  month        = dec,
  number       = {LBNL-2001637},
  doi          = {10.71468/P1WC7Q},
  address      = {Berkeley, CA}
}

@article{lublin2003parallel,
  author  = {Lublin, Uri and Feitelson, Dror G.},
  title   = {The Workload on Parallel Supercomputers: Modeling the Characteristics of Rigid Jobs},
  journal = {Journal of Parallel and Distributed Computing},
  volume  = {63},
  number  = {11},
  pages   = {1105--1122},
  year    = {2003},
  doi     = {10.1016/S0743-7315(03)00108-4}
}

@book{feitelson2015workload,
  author    = {Feitelson, Dror G.},
  title     = {Workload Modeling for Computer Systems Performance Evaluation},
  publisher = {Cambridge University Press},
  year      = {2015}
}

@article{gneiting2007scoring,
  author  = {Gneiting, Tilmann and Raftery, Adrian E.},
  title   = {Strictly Proper Scoring Rules, Prediction, and Estimation},
  journal = {Journal of the American Statistical Association},
  volume  = {102},
  number  = {477},
  pages   = {359--378},
  year    = {2007},
  doi     = {10.1198/016214506000001437},
  url     = {https://doi.org/10.1198/016214506000001437}
}

@article{higham2002nearestPSD,
  author  = {Higham, Nicholas J.},
  title   = {Computing the Nearest Correlation Matrix---A Problem from Finance},
  journal = {IMA J. Numerical Analysis},
  volume  = {22},
  number  = {3},
  pages   = {329--343},
  year    = {2002}
}

@article{bibbona2008ornstein,
  title   = {The {Ornstein--Uhlenbeck} process as a model of a low pass filtered white noise},
  author  = {Bibbona, Enrico and Panfilo, Gianna and Tavella, Patrizia},
  journal = {Metrologia},
  volume  = {45},
  number  = {6},
  pages   = {S117--S126},
  year    = {2008}
}

@article{sun2024energy,
  title     = {Energy dataset of {Frontier} supercomputer for waste heat recovery},
  author    = {Sun, Jian and Gao, Zhiming and Grant, David and Nawaz, Kashif and Wang, Pengtao and Yang, Cheng-Min and Boudreaux, Philip and Kowalski, Stephen and Huff, Shean},
  journal   = {Scientific Data},
  volume    = {11},
  number    = {1},
  pages     = {1077},
  year      = {2024}
}

@techreport{nerc2025largeloads,
  author      = {{North American Electric Reliability Corporation}},
  title       = {Characteristics and Risks of Emerging Large Loads: Large Loads Task Force White Paper},
  institution = {NERC},
  month       = jul,
  year        = {2025}
}

@techreport{bessasparis2025forecasting,
  title       = {Forecasting Large Loads in the Age of {AI} and Data Centers},
  author      = {Bessasparis, Stephen and Ramirez, Shana and Zhao, Emily and Zheng, Ye and Patel, Kush and Riu, Isabelle},
  institution = {{Energy and Environmental Economics, Inc. (E3)}},
  type        = {White paper},
  month       = dec,
  year        = {2025}
}

@inproceedings{Chaudhary2025DataCenterStability,
  author    = {Chaudhary, C. and Abdelkader, A. and Egan, M. and Udren, E. and Benidris, M. and Mitra, J.},
  title     = {Impact of Data Center Load Modeling on Power System Stability},
  booktitle = {Grid of the Future Symposium, CIGRE US},
  year      = {2025},
  month     = nov,
  address   = {Denver, CO, USA}
}

@inproceedings{chaudhary2026pesgm,
  author    = {Chaudhary, Chandan and Abdelkader, Alaaeldein and Pei, Yansong and Benidris, Mohammed and Mitra, Joydeep},
  title     = {Spatial Load Correlation in {AI} Data-Center-Dominated Power Systems},
  booktitle = {2026 IEEE Power \& Energy Society General Meeting (PES GM)},
  address   = {Montr{\'e}al, QC, Canada},
  month     = jul,
  year      = {2026}
}

@inproceedings{chaudhary2026pmaps_adequacy,
  author    = {Chaudhary, Chandan and Abdelkader, Alaaeldein and Benidris, Mohammed and Mitra, Joydeep},
  title     = {Resource Adequacy Risk in Correlated Large Loads},
  booktitle = {Int. Conf. Probabilistic Methods Applied to Power Systems (PMAPS)},
  address   = {Salt Lake City, UT, USA},
  year      = {2026}
}

@inproceedings{Chaudhary2026Modal,
  author    = {Chaudhary, C. and Murillo, M. and Benidris, M. and Mitra, J. and Pandit, D. and Bera, A.},
  title     = {Modal Analysis of Spatial Load Correlation in {AI} Data Center-Dominated Power Systems},
  booktitle = {Proc. IEEE Int. Conf. Smart Energy Systems and Technologies (SEST)},
  year      = {2026}
}

@misc{yan2026switching,
  title         = {Switching-Reference Voltage Control for Distribution Systems with {AI}-Training Data Centers},
  author        = {Yan, Mingyuan and Joswig-Jones, Trager and Zhang, Baosen and Chen, Yize and Cui, Wenqi},
  year          = {2026},
  month         = mar,
  eprint        = {2603.15588},
  archivePrefix = {arXiv},
  primaryClass  = {eess.SY},
  note          = {arXiv:2603.15588; GPU power traces released under the MIT license at github.com/yan-mingyuan/switchref}
}
    \endgroup

\end{document}